\documentclass[11pt]{article}
\pdfoutput=1

\usepackage{bm}
\usepackage{mathptmx}
\usepackage{etoolbox}
\usepackage{xcolor}
\usepackage{physics}

\usepackage[a4paper, margin=0.9in]{geometry}
\usepackage{authblk}
\usepackage{graphicx}
\usepackage{dcolumn}
\usepackage{amsmath}
\usepackage{amssymb}
\usepackage{textcomp}
\usepackage{verbatim}
\usepackage{float}
\usepackage[symbol]{footmisc}
\setfnsymbol{wiley}
\usepackage{hyperref}
\hypersetup{
    colorlinks=true,
    linkcolor=blue,
    filecolor=blue,
    urlcolor=blue,
    citecolor = blue,
    pdfauthor=author,
}

\usepackage{multirow}
\usepackage[T1]{fontenc}
\usepackage[normalem]{ulem}
\usepackage[square,sort&compress,comma,numbers]{natbib}

\begin{document}

\title{Perspectives on high-frequency nanomechanics, nanoacoustics, and nanophononics}

\author[1]{Priya}

\author[1]{E. R. Cardozo de Oliveira}

\author[1]{N. D. Lanzillotti-Kimura\footnote{daniel.kimura@c2n.upsaclay.fr}}

\affil[1]{Université Paris-Saclay, CNRS, Centre de Nanosciences et de Nanotechnologies, 91120 Palaiseau, France}

\date{}
\maketitle
\begin{abstract}

Nanomechanics, nanoacoustics, and nanophononics refer to the engineering of acoustic phonons and elastic waves at the nanoscale and their interactions with other excitations such as magnons, electrons, and photons.  This engineering enables the manipulation and control of solid-state properties that depend on the relative positions of atoms in a lattice. The access to advanced nanofabrication and novel characterization techniques enabled a fast development of the fields over the last decade. The applications of nanophononics include thermal management, ultrafast data processing, simulation, sensing, and the development of quantum technologies.  In this review, we cover some of the milestones and breakthroughs, and identify promising pathways of these emerging fields.
\end{abstract}

\maketitle 

\section{Introduction}
Usually seen as a major source of decoherence, acoustic phonons recently became an asset for optoelectronics~\cite{zarifi_highly_2018,stiller_coherently_2020,dainese_stimulated_2006}, quantum technologies~\cite{chafatinos_polariton-driven_2020}, and the simulation of solid-state systems~\cite{ortiz_topological_2021}. Indeed, all the properties of a solid-state structure that are determined by the position of the atoms are subject to be modified or modulated by the presence of acoustic phonons. Moreover, acoustic phonons can be used to interface platforms based on different excitations (electrons, excitons, photons, or magnons) through the engineering of their interactions.

The engineering of acoustic waves at the nanoscale covers a broad range of frequencies from MHz (ultrasound) to GHz (hypersound) and THz (heat).  Depending on the scientific community, it is refereed as \textit{nanomechanics}~\cite{bachtold_mesoscopic_2022} (i.e., when referring to NEMS and MEMS, acronyms for nano- and microelectromechanical systems, respectively), \textit{nanoacoustics}~\cite{maccabe_nano-acoustic_2020} (i.e., in cavity optomechanics), or \textit{nanophononics}~\cite{balandin_nanophononics_2005,lanzillotti-kimura_nanowave_2006} (i.e., in spectroscopy, hypersound engineering, and thermal transport). The focus of this review is to highlight the latest advances and trends and present some of the open challenges and future perspectives in the field.

By engineering nanostructures it is possible to control the confinement and propagation of acoustic vibrations, unlocking a plethora of applications.  The short wavelength (in the 1-100 nm range) associated with acoustic phonons of GHz-THz frequencies opens the possibility of both developing novel acoustic nanoscopies based on acoustic nanowaves, and reaching ultrasmall interaction mode volumes necessary for the future on-chip integrated phononic devices. Acoustic phonons constitute an efficient and versatile simulation platform. The long mean free path combined with the short wavelength enables the simulation of complex solid-state systems. In addition, one can have access to the full phonon wavefunction, which is difficult or impossible using platforms based on electrons and photons. The characteristic high frequencies in nanophononics have associated low occupation numbers, reaching the quantum regime using standard cryogenic techniques, paving the way for quantum applications where phonons play a central role to manipulate information.

Despite all these promising perspectives, nanophononics still faces some open challenges. The lack of standard transducers, in opposition to classical acoustics, motivates the development of novel strategies and experimental techniques to control the interactions with other excitations. A stringent requirement of nanophononic devices is the materials and the superior interface quality. With required roughness down to atomic scales for the higher frequency bands, it is the most limiting property for raising the working frequency. By overcoming these challenges, one will be able to address the full control of interaction mechanisms in order to efficiently generate, manipulate the dynamics, and detect acoustic phonons.

\begin{figure}
\begin{center}
\includegraphics[scale = 0.4]{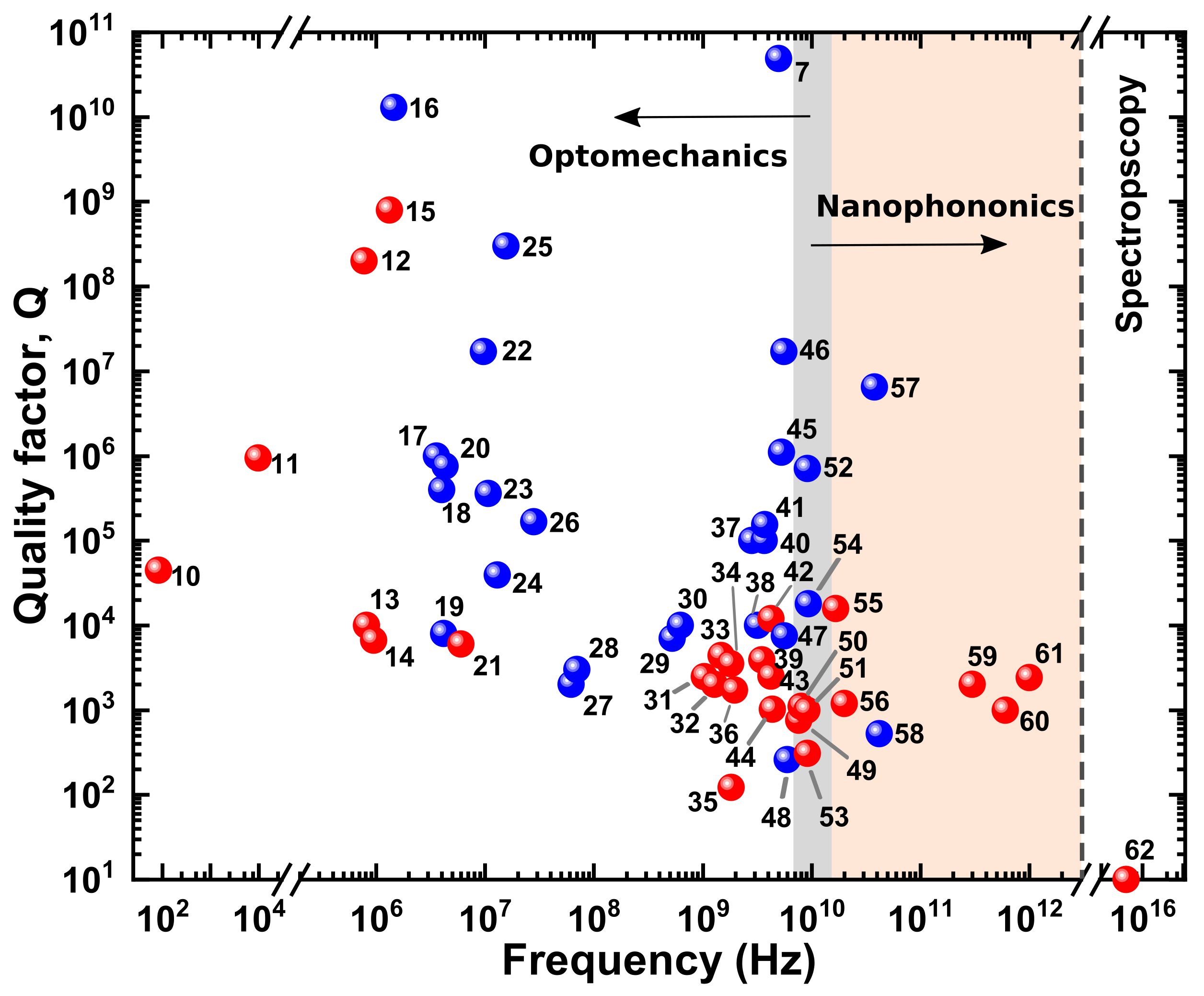}
\end{center}
\caption{\label{fig1} The acoustic quality factor (symbols) plotted as function of designed acoustic frequency for an assembly of acoustic systems operating at cryogenic (blue) and room (red) temperatures~\cite{mow-lowry_cooling_2008,kleckner_optomechanical_2011,tsaturyan_ultracoherent_2017,arcizet_experimental_2008,groblacher_observation_2009,ghadimi_elastic_2018,beccari_strained_2022,wollman_quantum_2015,weinstein_observation_2014,fink_quantum_2016,wilson_measurement-based_2015,butsch_cw-pumped_2014,dieterle_superconducting_2016,teufel_circuit_2011,pirkkalainen_squeezing_2015,goryachev_extremely_2012,safavi-naeini_squeezed_2013,schliesser_resolved-sideband_2009,riviere_optomechanical_2011,manenti_circuit_2017,liu_electromagnetically_2013,xiong_cavity_2013,shin_tailorable_2013,sun_high-q_2012,cleland_superconducting_2004,kang_tightly_2009,sun_superhigh-frequency_2012,schmidt_acoustic_2020,fan_aluminum_2013,grutter_slot-mode_2015,chan_laser_2011,krause_nonlinear_2015,sletten_resolving_2019,bochmann_nanomechanical_2013,kittlaus_large_2016,maccabe_nano-acoustic_2020,riedinger_non-classical_2016,meenehan_pulsed_2015,kervinen_sideband_2020,oconnell_quantum_2010,choudhary_advanced_2017,laer_thermal_2017,laer_net_2015,chu_quantum_2017,van_laer_interaction_2015,safavi-naeini_two-dimensional_2014,yaremkevich_protected_2021,fainstein_strong_2013,kharel_ultra-high-q_2018,kobecki_giant_2022,esmann_brillouin_2019,lanzillotti-kimura_resonant_2009,rozas_lifetime_2009,tarrago_velez_bell_2020}. The gray-shaded region represents the intersection of the fields of optomechanics (on the left) and nanophononics (orange-shaded region,on the right)}
\end{figure}

Figure~\ref{fig1} highlights the diversity of acoustic structures developed over the past two decades for the confinement of phonons in a cavity at different frequencies~\cite{mow-lowry_cooling_2008,kleckner_optomechanical_2011,tsaturyan_ultracoherent_2017,arcizet_experimental_2008,groblacher_observation_2009,ghadimi_elastic_2018,beccari_strained_2022,wollman_quantum_2015,weinstein_observation_2014,fink_quantum_2016,wilson_measurement-based_2015,butsch_cw-pumped_2014,dieterle_superconducting_2016,teufel_circuit_2011,pirkkalainen_squeezing_2015,goryachev_extremely_2012,safavi-naeini_squeezed_2013,schliesser_resolved-sideband_2009,riviere_optomechanical_2011,manenti_circuit_2017,liu_electromagnetically_2013,xiong_cavity_2013,shin_tailorable_2013,sun_high-q_2012,cleland_superconducting_2004,kang_tightly_2009,sun_superhigh-frequency_2012,schmidt_acoustic_2020,fan_aluminum_2013,grutter_slot-mode_2015,chan_laser_2011,krause_nonlinear_2015,sletten_resolving_2019,bochmann_nanomechanical_2013,kittlaus_large_2016,maccabe_nano-acoustic_2020,riedinger_non-classical_2016,meenehan_pulsed_2015,kervinen_sideband_2020,oconnell_quantum_2010,choudhary_advanced_2017,laer_thermal_2017,laer_net_2015,chu_quantum_2017,van_laer_interaction_2015,safavi-naeini_two-dimensional_2014,yaremkevich_protected_2021,fainstein_strong_2013,kharel_ultra-high-q_2018,kobecki_giant_2022,esmann_brillouin_2019,lanzillotti-kimura_resonant_2009,rozas_lifetime_2009,tarrago_velez_bell_2020}. This limited selection of structures is characterized by the acoustic quality factor Q. Q (symbols) are presented in Fig.~\ref{fig1} as a function of operating acoustic frequency. 
It is worth noting that acoustic structures with higher Q values at lower frequencies (MHz range) work at room temperature, while at higher frequencies (GHz range), cryogenic temperatures of a few mK are required. There is a demand for raising the Q-factor for room environment conditions. Even though this figure is not a detailed picture of the field, it is worth noting the proportion of works studying structures designed for the range of frequencies between tens to hundreds of gigahertz (highlighted in orange) is considerably smaller than at lower frequencies. Clearly, the development of high-Q resonators operating at high-frequencies, and able to work at room temperature, remains an open challenge in nanophononics.

The recent advances in material science, experimental techniques, nanofabrication methods, and the understanding of solid-state systems set the perfect framework and timing to explore the flourishing field of nanophononics~\cite{cang2022fundamentals,doi:10.1063/5.0042337}. The physics associated with acoustic phonons has implications in research fields, such as optomechanics, thermal nanophononics, bulk acoustic waves (BAWs), and surface acoustic waves (SAWs), which are closely connected to the topic of research discussed in this review. However, the operating frequency range, experimental techniques, and fabrication methods are different. The reader interested in thermal phonons, BAWs or SAWs is invited to refer to the respective review articles~\cite{zhang_coherent_2021,vasileiadis_phonon_2022-1,nomura_review_2022,liu_materials_2020,delsing_2019_2019,ng2022excitation}. In this article we will highlight some recent results in the fields of nanoacoustics and nanophononics, and then we will present what we believe are promising and exciting new avenues for the following years.

\section{Recent developments in nanophononics and nanoacoustics}
\subsection{Confinement of light and acoustic phonons}
The control of the acoustic nanowave dynamics is usually performed using acoustic resonators at the nano- and microscale. At the intersection between optomechanics and nanophononics, the state of the art is based on microdisks~\cite{liu_electromagnetically_2013,sun_high-q_2012,ding_wavelength-sized_2011}, microcavities~\cite{anguiano_micropillar_2017,ortiz_topological_2021}, nanobeams~\cite{maccabe_nano-acoustic_2020,navarro-urrios_room-temperature_2022}, and membrane-based resonators~\cite{tsaturyan_ultracoherent_2017,oudich2022tailoring}. Recent applications include sensing ~\cite{gil-santos_optomechanical_2020}, phonon lasing~\cite{chafatinos_polariton-driven_2020}, and the generation of non-classical phononic states~\cite{riedinger_non-classical_2016}. 

The fabrication of these devices relies on a top-down approach, thereby limiting the smallest achievable feature, and hence the working frequency range, generally to less than 10 GHz (See Fig.~\ref{fig1} and references therein). To reach even higher bands, bottom-up techniques warranting atomic-flat interfaces and nm-thick layers, such as molecular beam epitaxy and high-quality sputtering, are needed. Using thin films, multilayers, distributed Bragg reflectors, and Fabry-Perot resonators working up to 1 THz unlocked the domain of ultrahigh frequency nanophononics~\cite{lanzillotti2011coherent,levchuk_coherent_2020}. Fabry-Perot GaAs/AlAs-based nanocavities can be designed for the simultaneous confinement and colocalization of light and hypersound due to the equal ratios of speeds and impedances for light and sound in these materials~\cite{ortiz_topological_2021,anguiano_micropillar_2017, arregui_anderson_2019}. These resonators constitute an innovative platform for accessing phonon dynamics~\cite{lanzillotti-kimura_enhanced_2011} and cavity optomechanics applications.

\begin{figure*}
\includegraphics[scale = 0.5]{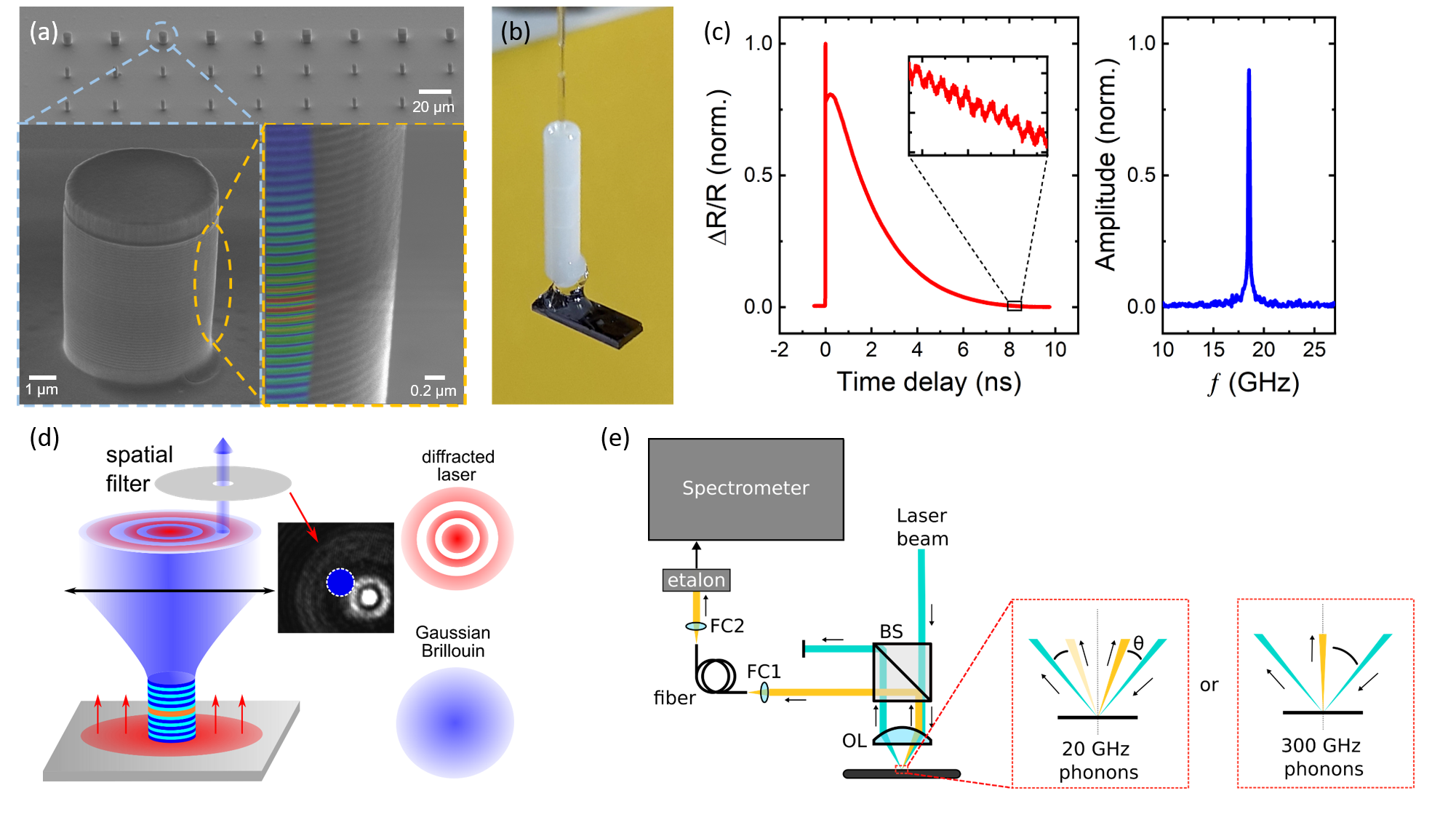}
\caption{\label{fig2} (a) SEM images of GaAs/AlAs micropillar cavities. Top panel: Array of etched micropillars of different sizes and shapes. Bottom left panel: Zoomed image of a single micropillar. Bottom right panel: Colormap of the displacement field highlighting the confinement of acoustic phonons in the cavity, superposed on the enlarged microscopy image. (b) Photo of a micropillar glued and mode-matched to a single-mode fiber. (c) Left panel: Reflectivity timetrace of a pump-probe experiment. An enlargement of the timetrace is shown on the inset, unveiling the presence of coherent oscillations. Right panel: FFT spectra of the displayed timetrace. (d) Interferometric-based filtering of acoustic modes of a 3D micropillar. (e) Angular filtering technique to access 20 GHz acoustic modes in Brillouin scattering for planar cavities.}
\end{figure*}

These multilayered cavities can be laterally etched to create optophononic micropillar cavities, as shown in Fig.~\ref{fig2}(a). This results in the 3D confinement of phonons and photons, thus tailoring the acoustic dispersion and local acoustic density of states~\cite{oudich2022tailoring}, as well as the optophononic interactions~\cite{anguiano_micropillar_2017, ortiz_topological_2021}. Semiconductor-based micropillar cavities are widely used in applications involving quantum dots~\cite{somaschi_near-optimal_2016}, quantum wells~\cite{bajoni_polariton_2008}, and optoelectronic devices, which makes them the perfect scenario for studying phononic interactions.

\subsection{Simulation of solid-state systems}
Analogous to band structures for photons and electrons in solid-state physics, one can engineer the acoustic band structure using multilayered periodic systems, such as superlattices (SLs) or coupled cavity systems. In these systems, it is possible to slow down a wavepacket, change the sign of propagation, and manipulate different mode symmetries~\cite{ortiz_phonon_2019}. Extending the electronic and acoustic bands analogy, it is possible to mimic any arbitrary potential. For instance, the Bloch oscillations of acoustic phonons~\cite{lanzillotti-kimura_bloch_2010} and topological interface states~\cite{ortiz_topological_2021} were implemented in both SLs-based and coupled-cavity systems~\cite{ortiz_phonon_2019,esmann_topological_2018}. The acoustic-phonon band engineering in SLs ~\cite{ortiz_phonon_2019} and coupled cavity structures~\cite{esmann_topological_2018} enable a versatile implementation of 1D tight binding and nearly free electron models using high-frequency phonons.

\subsection{Experimental techniques to access acoustic phonons}
The absence of standard characterization techniques for high-frequency acoustic phonons motivated the development of optical techniques both in the frequency and temporal domains. Picosecond ultrasonics based on pump-probe schemes using ultrafast lasers unveils the coherent phonon dynamics~\cite{thomsen_coherent_1984}, while high-resolution Raman and Brillouin scattering spectroscopy techniques unveil the incoherent phonon characteristics.  

The pump-probe technique allows the investigation of the reflectivity timetrace after a perturbation induced by an ultrashort pulsed laser (Fig.~\ref{fig2}c). In the picosecond ultrasonics framework, the coherent acoustic phonons, generated via photoinduced stress by the pump pulse~\cite{ruello_physical_2015-1}, modulate the refractive index and shape of the material. The coherently generated phononic signal is then imprinted on the transient optical reflectivity or transmission sensed by the probe pulse, and can be reconstructed by scanning the relative time delay between pump and probe pulses~\cite{thomsen_coherent_1984}. Depending on the lifetime of the acoustic phonons, the use of long optomechanical delay lines ($\sim 4$ m) is needed, which affects the stability of the spatiotemporal overlap of the pump and probe laser. This issue can be addressed by employing the asynchronous optical sampling (ASOPS) technique based on the traditional pump–probe experiment, using two mode-locked lasers pulsing at slightly different repetition rates~\cite{bartels_femtosecond_2006}. This technique eliminates the need for a mechanical delay line, thus increasing the stability of the setup. On the other hand, ASOPS requires two oscillators and the precise control on the repetition rates, which might be challenging. To deal with the mechanical stability issue differently, as well as the reproducibility of focalization, and optical mode matching, we recently developed fiber-integrated planar and micropillar optophonic cavities (Fig.~\ref{fig2}(b)) for pump-probe experiments~\cite{ortiz_fiber-integrated_2020}. In this approach, a single-mode fiber is aligned and glued onto an optophononic microcavity (planar or micropillar) with sufficient optical mode overlap and requires no further alignment steps in real-time experiments. This monolithic sample structure is potentially transportable to perform reproducible plug-and-play pump-probe experiments in individual microstructures at any laboratory.

Non-invasive Raman and Brillouin spectroscopy experiments based on the inelastic scattering of light from excited incoherent phonons are routinely performed to analyze the phononic modes of nanostructures~\cite{kargar_advances_2021}. In Raman scattering, the scattered signal from optical phonons (in THz range) is separated by a large frequency shift from the excitation laser, while reading out the small frequency shift of Brillouin scattered signal from acoustic phonons (GHz range) is a non-trivial task. Brillouin spectroscopy experiments usually incorporate multiple passage tandem interferometers or single-pass Fabry-Pérot spectrometers to record the Brillouin spectra of high-frequency acoustic phonons ranging from hundreds of MHz to GHz~\cite{rozas2014fabry,scarponi2017high,coker2018assessing,antonacci_breaking_2016}. More recently, in order to probe the Brillouin scattering spectra from phononic nanostructures and devices working at broad acoustic and optical wavelength ranges, we developed custom-built versatile experimental techniques  as sketched in Figs.~\ref{fig2}(d) and ~\ref{fig2}(e). Figure~\ref{fig2}(d) displays the diffraction-based spatial filtering method to assess the Brillouin scattering at 300 GHz in a 3D micropillar resonator~\cite{esmann_brillouin_2019}. In this scheme, mismatched spatial patterns of the diffracted laser excitation (shown in red) and the Gaussian-shaped Brillouin signal (shown in blue) are exploited to selectively collect the Brillouin signal transmitted through the node of the pattern. A fiber-based angular filtering technique is shown in Fig.~\ref{fig2}(e) to measure the confined acoustic mode in the range of 20-300 GHz in multilayered optophononic planar cavities~\cite{rodriguez_fiber-based_2021}. This scheme is based on the excitation (laser indicated in green) and collection of the enhanced Brillouin scattering signals (shown in yellow) through the optical cavity satisfying the double optical resonance condition ~\cite{fainstein1995raman}. The technique permits for better angular filtering of the signal using a single-mode fiber and employs etalon-filtering to further achieve high-spectral resolution at low-frequency phonons.

Noise spectroscopy, a conventional experimental technique in optomechanics, finds its limits in frequency, wavelength, or required sensitivity, motivating new developments in the fields at the intersection between cavity optomechanics and Brillouin spectroscopy. Electrical and magnetic generation of coherent acoustic phonons~\cite{maryam_dynamics_2013,yaremkevich_protected_2021} are challenging but promising research perspectives. Even the use of phase-changing materials (VO$_{2}$) as innovative transducers has been reported~\cite{mogunov_large_2020}.

\subsection{Novel nanoacoustic materials and structures}
Traditionally, acoustic phonon engineering is performed in highly crystalline and ordered systems, based on semiconductors, metals, and oxides systems. However, recent works demonstrate, on one hand, the feasibility and potential of nonperiodic systems such as disordered superlattices~\cite{arregui_anderson_2019}.  On the other hand, the use of easily available materials has been demonstrated in nanophononics that are based on reliable fabrication methods such as mesoporous~\cite{gomopoulos_one-dimensional_2010,schneider_engineering_2012,schneider_defect-controlled_2013,abdala_mesoporous_2020} medium, and polymer-based colloidal crystals~\cite{cheng_observation_2006,alonso-redondo_new_2015}. Acoustic frequencies in resonators based on mesoporous materials range between 5 and 40 GHz, with Q-factors in the order of $\sim$5-20~\cite{abdala_mesoporous_2020}. The chemical functionalization of resonators upon fluid adsorption finds its interest in sensing applications.~\cite{benetti_photoacoustic_2018} These functionalities and inexpensive fabrication processes are ideal for practical applications.

\subsection{Resonators for ultrahigh frequency phonons and light}
 The confinement of photons, phonons, and excitons in the same hybrid resonator provides a gateway to allow phononic control on the strong light-matter coupling regime~\cite{jusserand_polariton_2015,kobecki_giant_2022,chafatinos_polariton-driven_2020}. 

An optomechanical resonator facilitates the strong coherent coupling of photons to phonons resulting in the so-called dynamical back-action, which allows both the laser cooling~\cite{chan_laser_2011} down to the ground state of mechanical motion as well as the amplification of mechanical oscillations~\cite{kippenberg_cavity_2008}. A hybrid polariton-based optomechanical system was theoretically proposed to evince the cooling of mechanical motion at a single polariton level~\cite{restrepo_single-polariton_2014} and to reveal the unconventional dissipative coupling to mechanical mode~\cite{kyriienko_optomechanics_2014}. These hybrid systems are experimentally investigated to yield ultrastrong photoelastic coupling up to $10^{5}$ at exciton-polariton resonance in GaAs/AlAs multiple quantum wells~\cite{jusserand_polariton_2015}. This strong coupling of cavity polaritons to the high-frequency acoustic phonons in GaAs superlattices emanates as a prospective approach beneficial to experimentally achieve the cooling down of the mechanical vibrations to the quantum level. A recent study has introduced a new platform for the electrically driven longitudinal acoustic phonon-induced modulation of the polariton energies with amplitudes larger than the Rabi-splitting~\cite{kuznetsov_electrically_2021}. The electrically generated and detected coherent super high-frequency longitudinal bulk acoustic waves (LBAWs) have emerged as a powerful tool for governing different excitations in semiconductors~\cite{machado_generation_2019,crespo-poveda_ghz_2022} and provide an access to explore nonlinear acoustic processes such as second harmonic generation~\cite{bojahr_second_2015}, parametric oscillations and amplification featuring phonon lasing phenomenon.

A sound laser is characterized by the stimulated coherent emission of acoustic phonons and is also commonly referred to as a ‘phonon laser’~\cite{chafatinos_polariton-driven_2020,vahala_phonon_2009,otterstrom_silicon_2018,maryam_dynamics_2013}. The phonon lasing action has been theoretically advised and experimentally reported in several optomechanical systems at different acoustic frequencies ranging from MHz to hundreds of GHz~\cite{chafatinos_polariton-driven_2020,otterstrom_silicon_2018,maryam_dynamics_2013,grudinin_phonon_2010,cui_phonon_2021,mahboob_phonon_2013,lin_cascaded_2014}. At ultrahigh frequencies, the experimental observation of phonon amplification has been reported in hybrid systems coupling BEC with acoustic phonons~\cite{chafatinos_polariton-driven_2020}, and in semiconductor superlattices exploiting electron-phonon interactions~\cite{maryam_dynamics_2013}. 

\subsection{Nanoacoustics using optical nanoantennas}
Optical nanoantennas and, more recently, metasurfaces, revolutionized the field of optics. Their use in nanophononics came up as an original strategy to overcome one of the main roadblocks in the field, the lack of tailored and efficient transducers. Gigahertz mechanical modes are strongly dependent on the shape and materials of nanoantenna structures, and the interactions with light can be polarization-dependent, which enables a new playground for nanoacoustics. Indeed, single nanoantennas~\cite{berte_acoustic_2018}, or extended metasurfaces~\cite{obrien_ultrafast_2014,lanzillotti-kimura_polarization-controlled_2018,lanzillotti-kimura_control_2012} can be used as selective phonon generators, detectors and also for the demonstration of directional transport of phonons at the nanoscale~\cite{imade_gigahertz_2021,poblet_acoustic_2021}. Furthermore, plasmonic nanoparticles scaled down to 1 nm size range present acoustic resonances up to a few THz, which makes them potential acoustic terahertz generators~\cite{juve_probing_2010,crut_time-domain_2015}. An advanced engineering of the interaction of optical and GHz-acoustic fields in nanoantenna arrays is a promising arena for the development of optical, optoelectronic, and data communication applications.

\section{What are the perspectives on acoustic phonon engineering?}

\subsection{Tunable, responsive, and active optophononic devices}
\begin{figure}
\begin{center}
\includegraphics[scale = 0.4]{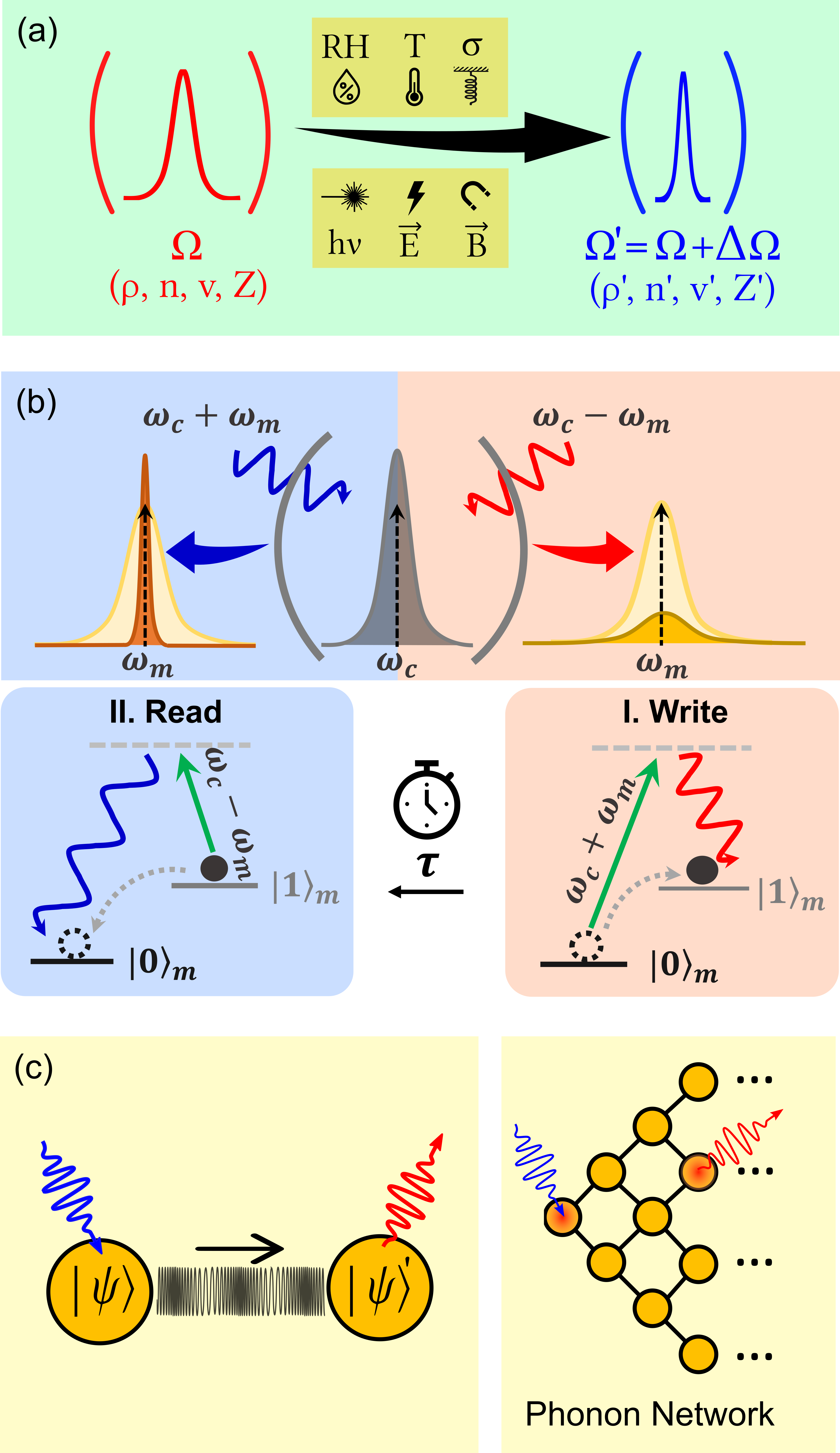}
\end{center}
\caption{\label{fig3} Schematics of perspectives on high-frequency acoustic phonons. (a) Tunable acoustic resonators: Under external environment control -- such as relative humidity (RH), temperature (T), strain ($\sigma$), energy ($h\nu$), electric field ($\vec{E}$) or magnetic field ($\vec{B}$)--, the acoustic resonator $\Omega$, that depends on the structural parameters mass density ($\rho$), index of refraction ($n$), speed of sound ($v$) and acoustic impedance ($Z$), undergoes changes on its resonance mode $\Omega' = \Omega + \Delta\Omega$ due to variation on the structure constants ($\rho'$, $n'$, $v'$,$Z'$). (b) Top panel: the principle of dynamical back action\cite{kippenberg_cavity_2008} applied to an optomechanical system operating in the resolved sideband regime where the frequency of the mechanical mode ($\omega_{m}$) is higher than the cavity line-width of optical mode ($\omega_{c}$). Upon interaction with the blue-detuned incident laser, there is an amplification of the mechanical mode, on the other hand, the red-detuned laser leads to the cooling of the mechanical mode. Bottom panel: Raman protocol for the creation (I. Write) and the annihilation (II. Read) of single phonon Fock state \cite{tarrago_velez_bell_2020} (c) Left panel: Interfacing different solid-state platforms via coherent acoustic phonons. The physical system $\ket{\psi}$ with engineered properties to couple with acoustic phonons, after excitation, transfers information to another system $\ket{\psi'}$ via coherent acoustic phonons, that can be sensed, e.g., optically. Right panel: Coupled acoustic resonators establishing a phonon network.}
\end{figure}

The development and integration of ultrahigh frequency optomechanical functionalities in Si-based platforms appear as a natural step in the evolution of the field.~\cite{zhang_subwavelength_2022} The functionalization and dynamical tuning of artificial acoustic materials, particularly at the nanoscale, would be a breakthrough in the engineering of intelligent and programmable materials. In nanoacoustics, despite the potential impact on data communications, sensing, optoelectronics, non-destructive testing, and in general on fundamental physics, with counted exceptions~\cite{alonso-redondo_new_2015}, no tunable nanophononic devices have been reported up to now. A reversible, and potentially ultrafast, change in the elastic properties (speed of sound, density, absorption, shape) of a nanostructure under external environment control (Fig.~\ref{fig3}(a)) would enable such dynamical behavior. Mesoporous thin films~\cite{abdala_mesoporous_2020}, susceptible of being functionalized and sensitive to environmental conditions, VO$_{2}$ showing a phase change associated with the emission of shock waves~\cite{mogunov_large_2020}, graphene and other 2D heterostructures having tunable electronic properties~\cite{zalalutdinov_acoustic_2021}, and magnetoelastic materials~\cite{kalashnikova_impulsive_2007,scherbakov_coherent_2010,thevenard_effect_2010,li_advances_2021} are possible ways towards the implementation of dynamical structures.

\subsection{Quantum nanoacoustics}
The core competencies for scalable quantum information processing are high bandwidth and long coherence times, which are fulfilled by coherent acoustic phonons. During the past few decades, the quantum nature of optomechanical systems has been investigated under two different experimental configurations as sketched in Fig.~\ref{fig3}(b): by creating a quantum ground state of mechanical motion under dynamical back-action (top panel)~\cite{kippenberg_cavity_2008} at cryogenic temperatures or using Raman scattering for creation and annihilation of a single phonon (bottom panel) in an ultrafast laser detection method~\cite{riedinger_non-classical_2016}. In recent experiments, optical phonons (40 THz)  in bulk diamond lattices are exploited to observe non-classical correlations~\cite{anderson_two-color_2018} and phonon entanglement~\cite{hou_quantum_2016}, and further to perform a Bell non-locality test~\cite{tarrago_velez_bell_2020}. Contrary to optical phonons, acoustic phonons are able to couple with various excitations (photons, electrons, etc) over a broad frequency range, and their interaction can be controlled at will in engineered structures for stronger interactions. For instance, the band structure engineering of acoustic phonons constitutes a unique asset. In this context, the emergent quantum technologies demand a quantum acoustical platform~\cite{wollack_quantum_2022,von_lupke_parity_2022,doi:10.1126/sciadv.add2811} operating at gigahertz frequencies providing single phonons for their use as quantum transducers, quantum sensors for displacement, and also as a probe in quantum thermodynamics.  

\subsection{Phonon transport and interfacing}
Owing to the strong coupling of acoustic phonons with other excitations, such as electrons, photons, excitons, plasmons, polaritons, and magnons, the interfacing of platforms based on different excitations is possible by engineering their interactions. Recently, the coupling of qubits with surface acoustic waves has been achieved~\cite{manenti_circuit_2017}. Furthermore, acoustic phonons can be employed as information carriers due to their long mean free path, which leads to potential application on signal processing and phononic networks (Fig.~\ref{fig3}(c)). However, previous works are frequency-limited up to a few GHz~\cite{imade_gigahertz_2021,poblet_acoustic_2021,stiller_coherently_2020,doi:10.1126/sciadv.add2811,kim_active_2021,florez_engineering_2022,mirhosseini_superconducting_2020}. By bringing these concepts to nanoacoustics and nanophononics it is possible to raise the frequency of the systems, thus reaching the quantum acoustic regime, and guide the quantum information as desired in complex acoustic networks. Inspired by developments in optics and electronics, topology appears as a promising tool towards the achievement of loose-less transport of nanophononic information and energy. In this sense, it is essential to develop and unlock 2D and 3D structures (resonators, waveguides, generators, and detectors) working in the GHz-THz regime where topological engineering can be defined.

\section{Conclusions}

We have shown our vision of the state of the art in phononic engineering at the tens to hundreds of GHz frequency range, highlighting exciting phenomena, such as the strong coupling with other excitations, and prospects in quantum applications. The nanoacoustics and nanophononics landscape is vast and difficult to cover in a short review. For instance, we did not cover the full field of bulk acoustic waves, high-impact applications like thermal transport management~\cite{cahill_nanoscale_2003}, nanoimaging and non-destructive testing, the study of biological tissues~\cite{margueritat_high-frequency_2019,dehoux_all-optical_2015,sandeep_-situ_2022,wang_imaging_2020}, or the potential of new research fields such as ultrahigh frequency sonochemistry~\cite{rezk_high_2021}. Combining these concepts with the ones presented in this article, incorporating novel materials, or bringing the acoustic resonators towards low occupation numbers will lead to a new generation of tunable and responsive nanophononic devices, quantum nanoacoustics, and the interfacing of different solid-state platforms via coherent acoustic phonons. The fast-growing fields of nanoacoustics and nanophononics will perhaps become the striking element for future quantum communication, sensing applications, and data communication technologies.

\section{Acknowledgements}
The authors acknowledge funding by the European Research Council Starting Grant No. 715939, Nanophennec, the French RENATECH network, and through a public grant overseen by the ANR as part of the “Investissements d’Avenir” program (Labex NanoSaclay Grant No. ANR-10-LABX-0035).

\bibliography{perspectives_nanophononics}

\end{document}